\documentstyle[12pt]{article}
\begin{document}
\setlength{\baselineskip}{2em}
\begin{titlepage}
\begin{center}
{\Large\bf Bose-Einstein Condensation May Occur in a Constant Magnetic
Field } \vskip 1.0 cm

{\bf H. Perez Rojas}$^{1,*}$\renewcommand{\thefootnote}{*}\footnote
{Permanent address: Grupo
de Fisica Te\'orica, ICIMAF,
Academia de Ciencias de Cuba, Calle E No. 309,Vedado, La Habana 4, Cuba.}
\vskip 1.5cm
$^1$ {\it Departamento de F\'{\i}sica}\\{\it Centro de
Investigaci\'on y de Estudios Avanzados del IPN}\\
{\it  Apartado Postal 14-740, 07000 M\'exico, D.F., M\'exico}

\end{center}

\begin{abstract}
\noindent
Bose-Einstein condensation of charged scalar and vector
particles may actually occur in presence
of a constant homogeneous magnetic field, but there is no
critical temperature at which condensation
starts. The condensate is described by the statistical distribution.
The Meissner effect is possible in the scalar, but not in the vector field
case, which exhibits a ferromagnetic behavior.
\end{abstract}
\noindent
\end{titlepage}

\newpage
It was pointed out long ago by Schafroth  \cite{Schafroth} that for a
non-relativistic boson gas, Bose-Einstein condensation (BEC) does not take
place  in the presence of a constant magnetic field. The
problem was studied afterwards by May \cite{May}, who
investigated the condensation in the non-relativistic case for an arbitrary
dimension $d$ and showed that it occurs for $d \geq 5$, and later by
Daicic {\it et al} \cite{Daicic},
\cite{Gailis}, who extended the considerations made by \cite{May} to the
relativistic high temperature case and found new magnetization properties.
More recently Toms \cite{Toms} has proved
that  BEC in presence of a constant magnetic field does not occur in any
number of spatial dimensions, and Elmfors {\it et al} \cite{Elm}
consider that a in the 3$d$ case, although a true condensate is not formed,
the Landau ground state can acommodate a large charge density. The
last paper discuss also the magnetization of the relativistic scalar gas
and shows that the Meissner effect  occurs in the low temperature case, in
analogy to the non-relativistic case studied by
\cite{Schafroth}. All these papers are characterized by very careful
calculations and their discrepancies, when arised, are in general related
to subtle points. One essential argument in all of them is
that condensation cannot occur since the condition for condensation, taken
from the symmetric (zero field) case as $\mu = \sqrt{ M c^2 \pm e B \hbar
c}$
for the relativistic case ($\mu' = \mu - M c^2 \pm e B \hbar/m c = 0$ in the
nonrelativistic limit), when applied to the case in which there is an
external magnetic field, leads to a divergent behaviour of the density in
terms of $\mu'$.

There are two different
ideas
which usually are considered to be the same, concerning what is to be
understood as BEC: 1) The existence
a finite fraction of the total
particle density in the ground state at some  temperature $T > 0$. 2)
The existence of a critical
temperature $T_c > 0$ for which $\mu' = 0$, such that for $T < T_c$ some
significant amount of particles start to condense in the ground state.

The present author's point of view is that i) the condition for
condensation in
the magnetic field case corresponds to 1), i.e. it {\it cannot}
be extrapolated as the same one of the standard zero field case, (which
is 2)), since in the magnetic field case there are different physical
conditions: explicit spatial symmetry breaking, discrete Landau states; ii)
since the chemical potential {\it is not} an independent thermodynamical
variable, but depends on the density, temperature and magnetic field, {\it
there is no divergence problem} at all, iii) as different from the zero
field case, in presence of the magnetic field the expresion for the density
contains the ground state contribution.

Let us remind the origin of the critical quantities
$\mu_c$, $T_c$ in the standard theory
of BEC (with zero magnetic field).
The chemical potential $\mu' = f(N, T)$ is a decreasing function of
temperature at
fixed density $N$, and for $\mu' = 0$ one gets an equation defining $T_c
=f_c (N)$. For temperatures $T < T_c$, as $\mu' = 0$, the
expression for
the density gives  values $N'(T) < N$, and the difference $N - N'= N_0$ is
interpreted as the density of particles in the condensate. This is to be
expected since the density of states, proportional to $4 \pi p^2$, cancels
the infrared divergence of the Bose-Einstein distribution $n(p, T)$ for
$\mu = 0$. The expression for $N$ contain only the particles in excited
states.
Thus, for $\mu'\to 0$, taking $p_0 \sim \sqrt{-2 M \mu'}$, the
amount of particles in a
neighbourhood of the ground state of amplitude $p_0$ is $N_0 =4 \pi
\int_0^{p_0} p^2 n(p, T)d p \sim M T p_0/\hbar^3$. In the magnetic field
case,
the infrared divergence of the Bose- Einstein distribution resulting from
taking $\mu' = 0$ is
not cancelled by the density of states in momentum space at $T \neq 0$.
In other words, for a given constant density, we are not allowed to put
$\mu' = 0$ keeping $T
\neq 0$. Now the amount of particles in a small neighbourhood of the
ground state of amplitude $p_{0}$, as we shall see below, is $N_0 \sim e B
\int_0^{p_0} n(p_3,
T)d p_3 \sim e B M^{1/2} T/\sqrt{-\mu'}$, which is large for small $\mu'$
and leads to
conclude that in the magnetic field case the expression for the density
accounts for the contribution of the population in all quantum states.

Thus, in the magnetic field case the problem of finding a
macroscopic ground
state density is not  conditioned to have
$\mu' = 0$.

We shall investigate in more  detail the problem of BEC
in
the physical $d = 3$ space,  at large densities and strong magnetic fields.
Under these conditions BEC occurs, but without having a definite critical
temperature $T_c > 0$, i.e. the phase transition is diffuse.
The concept of  diffuse phase
transitions, as those not having a definite critical temperature, but
an interval of $T$, was introduced long ago by
Smolenski and Isupov \cite{Smol} in their investigation of the phase
transition which occur in certain ferroelectric materials.

By assumming as in \cite{Elm} a constant microscopic magnetic field $B$
along the $p_3$ axis
(the external field is $H^{ext} = B - 4 \pi {\cal M}(B)$, where ${\cal
M}(B)$ is the magnetization),
the thermodynamic potential for a gas of scalar particles placed in it is

\begin{equation}
\Omega_s = \frac{eB}{4 \pi^2 \hbar^2 c \beta} \sum_{n = 0}^{\infty}
\int_{-\infty}^{\infty} dp_3 \left[\ln
(1 - e^{-(\epsilon_q - \mu)\beta})(1 - e^{-(\epsilon_q + \mu)\beta})
+ \beta \epsilon_q \right] ~\label{O}
\end{equation}
\noindent
where $\epsilon_q = \sqrt{ p_3^2 c^2 + M^2 c^4 + 2 eB \hbar c (n +
\frac{1}{2})}$, the last term in (1) accounts for the vacuum energy
and $\mu$ is the chemical potential. We shall use
$\beta = 1/T$, where $T = k t$ in what follows, $k$ being the Boltzmann
constant and $t$ the absolute temperature in Kelvins.

For a vector field the one-loop thermodynamic potential is
\begin{eqnarray}
\Omega_v& =& \frac{eB}{4 \pi^2 \hbar^2 c \beta}
\int_{-\infty}^{\infty} dp_3 \left[\ln
(1 - e^{-(\epsilon_{0q} - \mu)\beta})(1 - e^{-(\epsilon_{0q} +
\mu)\beta}) + \beta \epsilon_q \right] \nonumber
\\[1em]
&&\mbox{}
+ \frac{eB}{4 \pi^2 \hbar^2 c \beta} \sum_{n = 0}^{\infty}
\beta_n \int_{-\infty}^{\infty} dp_3 \left[\ln
(1 - e^{-(\epsilon_q - \mu)\beta})(1 - e^{-(\epsilon_q + \mu)\beta})
+ \beta \epsilon_q \right]
\end{eqnarray}
\noindent
where $\beta_n = 3 - \delta_{0n}$, $\epsilon_{0q} =
\sqrt{ p_3^2 c^2 + M^2 c^4 -  eB \hbar c}$, $\epsilon_q =
\sqrt{ p_3^2 c^2 + M^2 c^4 + 2 eB \hbar c (n + \frac{1}{2})}$.

The mean density of particles minus antiparticles (average charge
divided by $e$) is given by $ N_{s,v} = - \partial \Omega_{s, v}/
\partial \mu $.

We have explicitly
\begin{equation}
N_s = \frac{e B}{4 \pi^2 \hbar^2 c}\sum_0^{\infty}\left[\int^{\infty}_{-\infty}
dp_3 (n_p^+ - n_p^-) \right].
\end{equation}
\noindent
where $n_p^{\pm} = [exp(\epsilon_q \mp \mu)\beta - 1]^{-1}$.
For the vector field case we have,
\begin{eqnarray}
N_v & =&\frac{e B}{4 \pi^2 \hbar^2 c}\left[\int^{\infty}_{-\infty}
dp_3 (n_{0p}^+ - n_{0p}^-) \right]\nonumber
\\[1em]
&&\mbox{} + \frac{e B}{4
\pi^2 \hbar^2 c}\sum_0^{\infty}\beta_n \left[\int^{\infty}_{-\infty}
dp_3 (n_p^+ - n_p^-) \right]
\end{eqnarray}
\noindent
where $n_{0p}^{\pm} = [exp(\epsilon_{0q} \mp \mu)\beta - 1]^{-1}$,
$\epsilon_{0q} = \sqrt{ p_3^2 c^2 + M^2 c^4 - e B \hbar c}$, and
$\beta_{0n} = 3 - \delta_{0n}$.

For $T \to 0$, $\mu \to M_{\pm} c^2$  (we named
$M_{\pm} = \sqrt{M^2 \pm e B \hbar/c^3}$),
the population in Landau quantum states other than $n = 0$ vanishes (this
is shown explicitly in the Appendix) and the density  for the $n = 0$
state  is infrared divergent. We expect then most of the population to
be in the ground state, since for small temperatures $n_{0p}^-$ is
vanishing small and $n_{0p}^+$ is a bell-shaped curve with its maximum
at $p_3 = 0$. We will proceed as in \cite{Perez} and
call $p_0 (\gg \sqrt{- 2 M \mu'})$ some characteristic momentum.
We have then, by assumming $-\mu'
\ll T$, for the density in a small neighbourhood of $p_3 = 0$,

\begin{eqnarray}
N_{0 s,v} & = & \frac{e B T}{2 \pi^2 \hbar^2 c}\int_0^{p_0} \frac{d
p_3}{\sqrt{p_3^2 c^2 + M^2 c^4 \pm e B \hbar c} - \mu} \nonumber
\\[1em]
&&\mbox{}= \frac{e B T}{2 \pi^2 \hbar^2 c}\int_0^{p_0} \frac{d
p_3}{p_3^2/2M_{\pm} - \mu'} \nonumber
\\[1em]
&&\mbox{}= \frac{e B T}{4 \pi \hbar^2 c}\sqrt{\frac{2 M_{\pm}}{-
\mu'}}  \label{5}
\end{eqnarray}
\noindent
where $N_{s,v} = N_{0 s,v} + \delta N_{s,v}$ and $\delta N_{s,v}$ is the
density in the interval $[p_0, \infty]$. Actually $\delta N_{s,v}$ is
negligibly small and $N_{s,v}^0 \simeq
N_{s,v}$.

The expression (\ref{5}) was obtained in \cite{Perez} and more recently
in \cite{Elm}. Formally it indicates a divergent behaviour of $N_{0 s,v}$ for
$\mu'\to 0$, but we must be careful in doing that interpretation.
Actually it means nothing more that in a
neighbourhood of the ground state, at constant temperature,
$N_{0 s,v} \sim (\mu')^{-\frac{1}{2}}$, and for constant $N_{0 s,v}$, $\mu'
\sim T^2$. This comes from the fact that
$\mu'$ is not, and cannot be taken, as an independent thermodynamic
variable (see i.e. \cite{Landau}, chapter 3),
but depends on the quantities $N_{s,v}^0, T, B$ and it is enough that
$\mu'\sim T^2 f(N_{0 s,v})$ for $T \to 0$ ($f$ being a finite function) to
avoid the divergence.  Thus  (\ref{5}) simply means that
\noindent
\begin{equation}
\mu' = - \frac{e^2 B^2 T^2 M_{\pm}}{8 \pi^2
N_{0 s,v}^2 \hbar^4 c^2}. ~\label{3}
\end{equation}
We observe
that $\mu'$ is a decreasing function of $T$ and vanishes for $T = 0$,
where the "critical" condition $\mu = M_{\pm} c^2$ is reached.
As shown below, in that limit the Bose-Einstein
distribution degenerates in a Dirac $\delta$ function, which means to have
all the system in the ground state $p_3 = 0$. To see this, we shall
rewrite  the momentum density of particles around the
ground state $p_3 = 0$, $n = 0$ approximately as
\begin{eqnarray}
n_0 (p_3) &=& \frac{T}{\frac{p_3^2}{2 M_{\pm}} + \frac{e^2 B^2 T^2 M_{\pm}}
{8 \pi^2 \hbar^4 c^2 N_{0 s,v}^2}}\nonumber
\\[1em]
&&\hbox{}=\frac{4 \pi \hbar^2 c N_{0 s,v}}{e B}\cdot \frac{\gamma}{p_3^2 +
\gamma^2}
\end{eqnarray}
where
\[
\gamma = \frac{e B T M_{\pm}}{2 \pi \hbar^2 c N_{0 s,v}} =
\frac{p_T}{v^3 N_{0 s,v}} = \sqrt{- 2 M_{\pm} \mu'} \]
\noindent
where $p_T = \sqrt{2 \pi M_{\pm} T}$ is the thermal momentum, $v^3 =
h c\lambda/e B$ the elementary volume cell, $\lambda = h/p_T$ being
the De Broglie thermal wavelength. We have
approximated the Bose-Einstein distribution by one proportional to a
Cauchy distribution, having its maximum
$\sim \gamma^{-1}$ for $p_3 = 0$. We have that $\gamma \to 0 $ for $T \to 0 $,
but for small fixed $T$,
$\gamma$ also decreases as $ v^3 N_{0 s, v} $ increases. We remind that in
the zero
field case, the condensation condition demands $N \lambda^3 > 2.612$.

\noindent
One can write then
\begin{equation}
\frac{1}{2} N_{0 s,v} = \frac{ e B}{4 \pi^2 \hbar^2 c}
\int_{-\gamma}^{\gamma} n_0 (p_3) d p_3. \label{cond}
\end{equation}
Thus, approximately  one half of the total density
is concentrated in the narrow strip of width $2\gamma$ around the
$p_3 = 0$ momentum. It results that for densities and magnetic fields large
enough, if
we choose an arbitrary small neighbourhood of the ground state,
of momentum width $2 p_{30}$, one can always find a temperature
$T > 0$ small enough such that $\gamma \ll p_{30}$ and
(\ref{5}) and (\ref{cond}) are satisfied. The
condensate appears and it is described by the statistical distribution.

We have also

\begin{equation}
\lim_{\gamma \to 0} n_0 (p_3)  =  4 \pi^2 \hbar^2 c \frac{N_{s,v}}{e B}
\delta(p_3).
\end{equation}
and obviously
\begin{equation}
N_{s,v} = lim_{\gamma \to 0} \frac{ e B}{4 \pi^2 \hbar^2 c}
\int_{-\infty}^{\infty} n_0 (p_3) d p_3.
\end{equation}
\noindent
Thus, if $T = 0$, all the density $N_{s,v}$ lies in the
condensate, as occur in the zero field case, but here the total density
is described explicitly by the integral in momentum space.

Obviously, one cannot fix any  (small)  value for
$\gamma$ from which
the distribution starts to have a manifest  $\delta (p_3)$ behavior;
and {\it there is
no} critical
temperature for condensation to start, which we interpret as a diffuse phase
transition. But
when $\gamma$ and $1 / \Delta$ decrease enough (we define $\Delta =
\omega \hbar/ T$, and $\omega = e B/M c$. If $\Delta \ll 1$, the system
is confined to the $n = 0$ Landau
state, see below), the conditions for condensation mentioned
above are satisfied.

In the non-relativistic case, the above results can be derived even from
Schafroth's formulae. We
can write the non-relativistic limit for $N_s$ (which is a very good
approximation for
the relativistic case if $M c^2 \gg T$, since the main contribution to the
integrals comes from values of $p_3 \leq M c$ ), as
\begin{eqnarray}
N_s &= &\frac{e B}{2 \pi^2 \hbar^2 c}\sqrt{\frac{\pi M T}{2}} \sum_{m =
1}^{\infty}\frac{e^{-\mu_1 m \beta}}{m^{1/2}}\frac{1}{1 -
e^{- 2\omega \hbar m \beta}} \nonumber
\\[1em]
&&\hbox{}=\frac{e B}{2 \pi^2 \hbar^2 c}\sqrt{\frac{\pi M T}{2}} \sum_{m =
1}^{\infty}\frac{e^{-\mu_1 m \beta}}{m^{1/2}}\left[ 1 +
\frac{e^{- 2\omega \hbar m \beta} }{1 - e^{- 2\omega \hbar m \beta}}\right].
{}~\label{s}
\end{eqnarray}
where the unity in angular brackets accounts for the Landau $n = 0$ state
and the second term for all the set of excited states. We have used $\mu_1
= M c^2 + \omega
\hbar - \mu$ and $\omega = e B /M c$. Now, if the  parameter $\Delta =
\omega \hbar / T \gg 1$, and if $\mu_1  \to 0 $, even at large
temperatures, we can neglect the second term in angular brackets
in (\ref{s}), and approximate the resulting sum by an integral. This
is done
by introducing the
continuous variable $x = T m$ and by writing $T \sum_{n = 1}^{\infty} \to
\int_0^{\infty} dx$. After integrating, we obtain back (5), with
$\mu_1$, $M$ in place of $\mu'$, $M_{\pm}$.

Neutron stars, where
strong magnetic fields and very high densities are assumed to exist,
may provide conditions for the ocurrence of BEC. It has been conjectured
that superfluid and superconductive effects are produced (see i.e.
\cite{Thorsson}, \cite{Page} and references
therein).

As a simplified model of the gas of kaons in a neutron
star, we take  $ N_s \sim 10^{44}$ cm${}^{-3}$, $t
\simeq 10^8$${}^\circ$K,  and  local
magnetic fields
$B \sim 10^{14}$ G (fields of order $10^{15}$ G have been estimated inside
hadrons \cite{Linde}), the conditions for BEC can be
also satisfied,  if the medium provide also screening mechanisms for the
very large electric fields which also arise. In this case $\gamma =
10^{-30}$.
If we suppose the star of dimensions $\sim 10^7 {\rm cm}$, the discrete
momentum states would be spaced by an amount $\delta p \sim 10^{-34} {\rm g
cm/s}$. This means to have one half of the total density distributed in
these $2 \gamma/ \delta p = 10^4$ quantum states. The
ground state density, with strictly zero momentum $p_3$, can thus be
estimated as $10^{40}$ ${\rm cm}^{-3}$. We observe in this case the
dimensionless phase space density $N v^3 \sim 10^{12}$.

We turn now to the magnetization problem. From
(\ref{O}), we get that for $T \to 0$, the magnetization of the scalar
field is
\begin{eqnarray}
{\cal M}_s & =& -\frac{\partial \Omega}{\partial B} \nonumber
\\[2em]
&&\mbox{} = - \frac{e}{4 \pi^2 \hbar} \sum_0^{\infty}
\int_{-\infty}^{\infty}
d p_3 \frac{e (n + \frac{1}{2})}{\epsilon_q} (n_q^+ + n_q^-).
\end{eqnarray}
\noindent
In the condensation limit we get
\begin{equation}
{\cal M}_s = - \frac{e N_s \hbar}{2 M_+ c}. \label{m}
\end{equation}
\noindent

\noindent
We see that the magnetization is opposed to the external field and the
system behaves as a perfect {\it dia-ferromagnetic} (this was first pointed
out by Schafroth \cite{Schafroth}).

If we take the $B (> T)\to 0$ limit of (\ref{m}), one can discuss the
critical conditions for the arising of the Meissner effect in the low
temperature relativistic case. The condition (obtained  for $T \neq 0$ in
ref. \cite{Elm}), is $H_c^{ext} = -{\cal M}_s (0)$. In particular, we
agree with
\cite{Elm} that in the relativistic case the Meissner effect occurs in
analogy with the non-relativistic one, and that it is not
neccessarily connected
with high temperature pair creation proccesses, as suggested in
\cite{Daicic}.

For the vector field case, the
magnetization in the condensation limit is positive since all the system
is in the Landau $n = 0$ state, (see eq. (2)), and
\begin{equation}
{\cal M}_v = \frac{e N_v \hbar}{2 M_- c}
\end{equation}
\noindent
and we have that the condensate of vector particles behave as
a true ferromagnet. In particular, we can write $0 = H_c^{ext} = B - {\cal
M}(B)$ as the condition for spontaneous magnetization to occur.

In concluding, it is important to remind that for the first time true BEC
has been recently observed \cite{Anderson} in evaporated ${}^{87}Rb$ atoms
confined
in a magnetic trap creating an ellipsoidal potential. The system is far
from being an isotropic noninteracting gas, and although it also differs
from our present model, it has, however, some analogies.
In particular, the anisotropy of the confining potential leads to a
larger velocity spread in the axial direction as compared to the radial one,
which is
the analog of the $p_3$ spread of the $n(p_3)$ distribution in the magnetic
field case, confined radially to the $n =0$ Landau quantum state.

The author thanks R. Baquero, J. Hirsch, K. Kirsten, D. B. Lichtenberg, O.
Perez- Martinez and D. Quesada for  comments.
He is especially indebted to A. E. Shabad for a discussion long ago, from
which some of the  basic ideas presented in this paper arised.
\begin{center}
{\bf Appendix}
\end{center}
We will present a demonstration concerning the vanishing of the density for
excited Landau states if $\Delta \gg 1$, $M \beta \gg 1$. We use in this
section units $\hbar = c = 1$, and consider only the scalar field case.
Let us call $N_e$ the density corresponding to excited states. We can
write, by using the $K_2$ Bessel function representation in
terms of dimensionless quantities $\bar M_+ = M_+ \beta$, $\bar
M_+ = M_+ \beta$ and $\bar
\mu = \mu \beta$, after summing over Landau quantum numbers from 1 to
$\infty$,
\begin{equation}
N_e = \frac{e B}{8 \pi^2} \sum_{m = 1}^{\infty} m \sinh m \bar
\mu \int_0^{\infty}\frac{ dt}{t^2} e^{-\frac{m^2}{4 t} - \bar M_+^2 t} \cdot
\frac{e^{-2 e B \beta^2 t}}{ 1 - e^{-2 e B \beta^2 t}}.
\end{equation}

We will establish an upper bound to this quantity. Let us introduce $x =
\bar M_+ t$ and cut the integral in two parts by the point $x_0
= \theta/(\bar M_+ + \Delta)$, where $\theta \ll 1$. Let us
call also $\xi$ to some point in between $x_0$ and $\infty$. We have

\begin{equation}
N_e \leq \frac{M^2}{8 \pi^2 }\sum_{m = 1}^{\infty} m \sinh m
\bar \mu \left[ \frac{16}{m^4 \bar M^2} e^{-\frac{m^2 \bar M_+^2}{\theta}} +
\frac{2 \omega \hbar e^{-m^2 \bar M_+ /4
\xi}}{\xi^2}\cdot
\frac{e^{-2 \Delta \xi}}{1 - e^{-2 \Delta \xi}}\cdot \frac{e^{-(\bar
M_+) x_0}}{\bar M_+}\right]
\end{equation}

Both series can be made to converge  to an arbitrary small number as $\theta
\to 0$ and
 $1/\Delta \to 0$, provided that $\bar M \gg 1$.


\begin{thebibliography}{99}
\frenchspacing
\bibitem{Schafroth} M. R. Schafroth, {\it Phys. Rev.} {\bf 100} (1955)
463.
\bibitem{May} R. M. May, {\it J. Math. Phys.} {\bf 6} (1965) 1462.
\bibitem{Daicic} J. Daicic, N. E. Frankel, and V. Kowalenko, {\it Phys.
Rev. Lett.} {\bf 71} (1993) 1779.
\bibitem{Gailis}J. Daicic, N. E. Frankel, R. M. Gailis and V. Kowalenko,
{\it Phys. Rep.}{\bf 237} (1994) 63.
\bibitem{Toms} D. J.Toms,{\it Phys. Rev. Lett.} {\bf 69}
(1992), 1152; {\it Phys. Rev.} {\bf D 47} (1993), 2483;
{\it Phys. Lett.} {\bf B 343} (1995), 259
\bibitem{Elm} P. Elmfors, P. Liljenberg, David Persson, Bo-Sture Skagerstam,
{\it Phys. Lett.}{\bf B 348} (1995), 462.
\bibitem{Smol} G. A. Smolenski and V. A. Isupov, {\it Sov. Journal of
Techn. Phys.} {\bf 24} (1954) 1375.
\bibitem{Perez} H. Perez Rojas, {\it Acta Phys. Pol.} {\bf B 17} (1986), 861.
\bibitem{Landau} L. D. Landau and E. M. Lifshitz, {\it Statistical Physics},
3rd. Ed., Part 1, Pergamon Press (1986), New York.
\bibitem{Thorsson} V. Thorsson, M. Prakash and J. M. Latimer, {\it Nucl.
Phys.}{\bf A 572} (1994), 693.
\bibitem{Page} Dany Page and James H. Applegate, {\it The Astrophys.
Journal} {\bf 394} (1992) L 17.
\bibitem{Linde} A. D. Linde, {\it Rep. Prog. Phys.} {\bf 42} (1979), 389.
\bibitem{Anderson} M. H. Anderson, J. R. Ensher, M. R. Matthews, C. E.
Wieman, E. A. Cornell, {\it Science} {\bf 269} (1995), 198.
\end{thebibliography}
\end{document}